# Evaporation of tiny water aggregation on solid surfaces of different wetting properties

Shen Wang[1,2], Yusong Tu[3], Rongzheng Wan[1,*] and Haiping Fang[1]

[1] *Shanghai Institute of Applied Physics, Chinese Academy of Sciences, P.O. Box 800-204, Shanghai, 201800, China.*

[2] *Graduate School of the Chinese Academy of Science, Beijing 100080, China*

[3] *Institute of Systems Biology, Shanghai University, Shanghai, 200444, China.*

## Abstract

The evaporation of a tiny amount of water on the solid surface with different wettability has been studied by molecular dynamics simulations. We found that, as the surface changed from hydrophobicity to hydrophility, the evaporation speed did not show a monotonically decrease from intuition, but increased first, and then decreased after reached a maximum value. The competition between the number of the water molecules on the water-gas surface from where the water molecules can evaporate and the potential barrier to prevent those water molecules from evaporating results in the unexpected behavior of the evaporation. A theoretical model based on those two factors can fit the simulation data very well. This finding is helpful in understanding the evaporation on the biological surfaces, designing artificial surface of ultra fast water evaporating or preserving water in soil.

## Keywords:

# Introduction:

The evaporation of water is very important in biological[1] and environmental[2] science. The water evaporation of bulk surface such as a lake or a river is a classical topic and has been studied for long time[2-4]. Recent works show that the nanoscale confined water and water restricted on the surface of solid materials is ubiquitous in nature[5-15], e.g. the ice-like water monolayer on top of a hydrophilic[6,16,17] or a hydrophobic[18-21] solid surface. The evaporation of such nanoscale water aggregation is essential in the macroscopic world, e.g., the water loss through the surface of soil[22], and water aggregation above the surfactant[23-25]; and the evaporation of the surficial water aggregation also plays an important role in biophysical phenomena such as the folding of globular protein[26-29]. The evaporation of such confined water aggregation is different from the classical evaporation of bulk water. Very recently, Lee and his co-workers have used scanning electric microscope to study the evaporation efficiency of micro-scaled water droplet from nanoporous microcantilevers of various hydrophobicity and stated that the dynamics of water evaporation between hydrophobic and hydrophilic conditions are very different[30,31]. But how the surface wettability affects the evaporation of the tiny water aggregation still remains unclear.

Here we present molecular dynamics (MD) simulations on the evaporation of nanoscale water aggregation on a solid substrate with different surface wettability at room temperature. The water aggregation was adhered to a solid wall-liked substrate and the thickness of the water layer was less than 1 nanometer. The simulations showed that the evaporation speed of water aggregation first increased as the surface wettability increased when the surface wettability was not high enough; when the surface wettability was high enough, the evaporation speed of water aggregation decreased as the surface wettability increased. Hence, the surface wettability can affect the evaporation flux in two mechanisms: the increasing of surface wettability enhances the evaporation flux by increasing the area of liquid-gas surface of the water aggregation; the increasing of surface wettability reduces the evaporation flux by increasing the escape energy barrier of surficial water molecules. The final evaporation speed of nanoscale water aggregation is the competition of these two mechanisms. When the surface wettability is low, the surface wettability dominatingly increases the area of liquid-gas surface which causes the increasing of evaporation speed; after the surface wettability goes high enough, the surface wettability dominatingly increases the escaping energy barrier of surficial water molecules which causes the decreasing of the evaporation speed.

## Simulation Method:

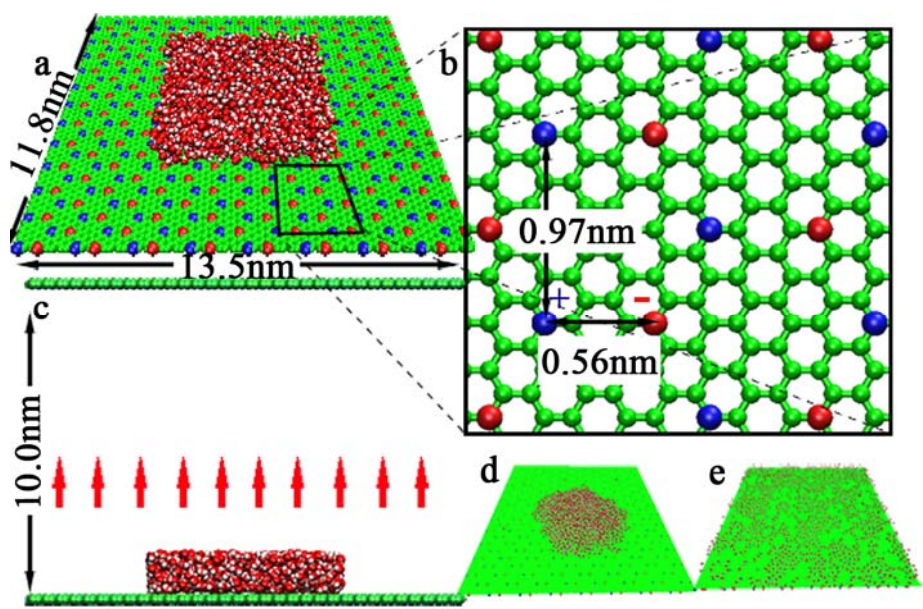

Figure 1(a) The initial system (b) Detail geometry of the solid substrate model. Red and blue spheres represent the atoms with positive and negative charges, respectively, while the green spheres represent neutral atoms. (c) Side view of the initial system. The upward arrow denoted the accelerating region. (d) Snapshot of water on the substrate with low wettability, $q = 0$ e. (e) Snapshot of water on the substrate with high wettability, $q = 0.7$ e.

MD simulations were carried out in a box with initial size of 13.5 nm × 11.8 nm × 11.0 nm. We used the Charmm22 force field with a rigid TIP3P[32] water model, the atoms in the graphite-liked model solid substrate had the Lennard-Jones parameters $\varepsilon$ = 0.07 kCal/mol, $\sigma$ = 0.4 nm which were slightly modified from the aromatic carbon[33,34]. The periodic boundary condition was applied to all three independent Cartesian coordinates x, y, z. The surface wettability of a substrate can be described by the contact angle of the spherical cap shaped sessile water droplet laying on it. However such definition of surface wettability is not absolute because the contact angle can also be affected by the amount of aqueous[35] or nearby liquid molecules[36].

Hence, although the contact angle θ is used as a phenomenology criterion of surface wettability of the substrate in this study, it may not be consistent to the experimental result of relatively larger droplet. Similar to the work of Wang et al[37], the substrate had dimensions of 13.5 nm × 11.8 nm with a planar hexagonal structure of neighboring bond lengths of 0.142 nm. As shown in Figure 1ab, positive and negative charges of the same magnitude *q* were assigned on the atoms at a distance of 0.568 nm (2 times of the diagonal of the hexagon). Overall, the substrate contained 192 positive charges and 192 negative charges, and was neutral. By changing the value of *q*, we obtained a substrate with tunable surface wettability which was inspired by the laboratory available polar surface[38-41]. This graphite-like sheet lay on the bottom of the cell along the x-y plane, above which the water monolayer was laid (Figure 1ac). There were totally 1298 water molecules in the system. The number of water molecules was carefully selected that the water molecules could form a perfect monolayer on the substrate when the wettability of the substrate was high enough. There was a ceiling solid wall with the same dimension of the substrate laying 10nm higher to prevent the water molecules from reaching the bottom side of the substrate. An accelerating region was applied from 3nm to 5nm above the substrate as shown in Figure1c. When a water molecule ran into the accelerating region, a vertically upward force as strong as 1 kCal×mol$^{-1}$A$^{-1}$ (force vector (0 0 1)) was applied on the oxygen atom of that water molecule to prevent the water molecule from going back to the substrate. Such nonequilibrium condition is equivalent to the setting of some other literature that the evaporated molecules will run into an infinite vacuum or reach a zero potential point (infinite far from the liquid)[42]. Thus, an obvious evaporation process can be observed during the limited 10 ns MD simulation without the influence of the atmosphere pressure[43].

Initially, the 1298 water molecules were piled upon the substrate in a water cube of 6 nm × 6 nm × 1.2 nm, as shown in Figure 1a. After an equilibrium MD simulation (without the accelerating region) for 10 nanoseconds, the water molecules formed a spherical shaped sessile droplet on the substrate with low wettability (Figure 1d) or a smoothly spread monolayer on the substrate with high wettability (Figure 1e). The surface wettability of the substrate could be denoted by the contact angle of the water droplet and the interaction energy between the water molecules and the substrate. Then a nonequilibrium MD simulation (with the accelerating region) was carried out to study the evaporation behavior of the water molecule on substrate with different wettability.

All the MD simulations were carried out by NAMD2.6[44] with a time step of 2fs, and the Particle Mesh Ewald (PME) integration was employed with a frequency of once per 4 steps. The PME Grid size in X, Y and Z direction were 1/128, 1/128 and 1/100 of the box size. All the water molecules were attached to a thermostat of 300 K and a simple method of Berendsen Temperature coupling[45] was applied every 20 steps, which was also the same frequency of atom reassignment. The trajectories of the coordinate and velocity were collected every 500 steps, i.e., every 1picoseconds, and totally 10000 frames of datum were included in the trajectory files.

## Result and discussion:

The simulations show that the water spreads smoothly on the substrate when $q \geq 0.4$ e; when $q < 0.4$ e, the water shrinks gradually into a sessile droplet like a spherical cap as $q$ decreased. The contact angle $\theta$ is used as a phenomenology criterion of surface wettability. In the simulation, $q$ is sampled from 0-0.7e with an increment of 0.1 e that both droplet and monolayer form of the water can be studied.

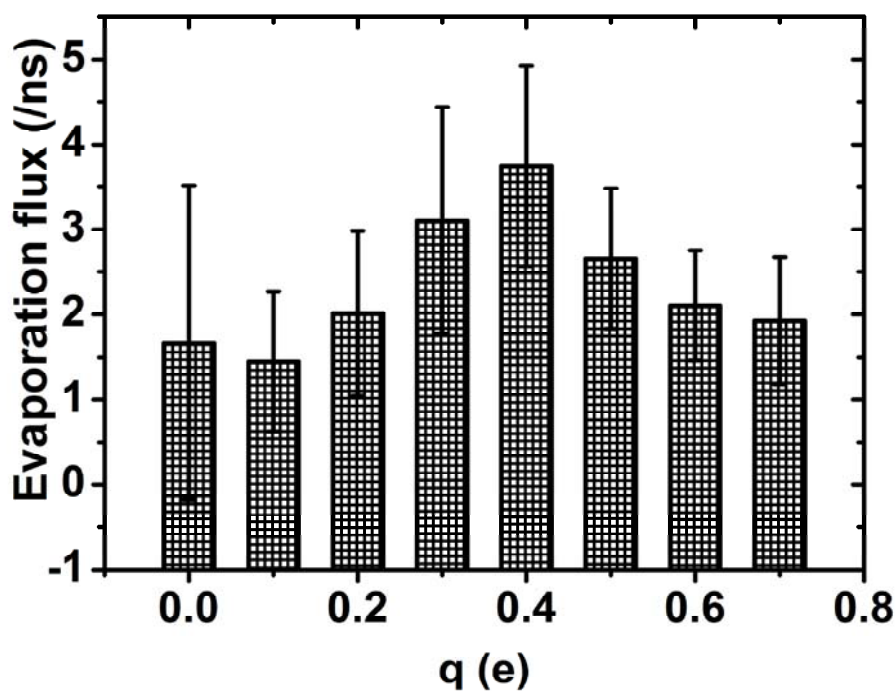

Figure 2 Evaporation flux and its error versus different assigned charge *q*.

The evaporation speed of the water layer can be described by the evaporation flux which is defined as the average number of the water molecules entering the accelerating region from the substrate per nanosecond, since such water molecules will go upward till reaches the ceiling and will not go back to the substrate as condensed molecules. Counter to intuition, the evaporation flux does not decrease monotonically as *q* increases. Actually, as shown in Figure 2, the evaporation flux first increases as *q* increases when *q* < 0.4 e; then the evaporation flux reaches its maximum around *q* = 0.4 e; when *q* ≥ 0.4 e, the evaporation flux decreases as *q* increases.

To analyze this unusual variation of the evaporation flux, we assume that the evaporation flux $J(q)$ can be denoted as a product of the probability for a water molecule on the liquid-gas surface of the water molecules aggregation and the escape probability of such surficial water molecule.

$$J(q) \propto P_{geo}(\theta(q)) P_{ener}(E) \qquad (1)$$

Here $P_{geo}(\theta)$ is the probability for a water molecule on the liquid-gas surface, which is a geometry related factor. When the water molecules escape from a sessile droplet on the substrate, only those on the liquid-gas surface region have the opportunity to run away. The probability for the water molecules on this outermost region varies with the shape of the water droplet, and can be denoted as a function of the contact angle θ of the water droplet. $P_{ener}(E)$ is the escape probability of a surficial water molecule. E = $E_{WW}$ + $E_{sub}(q)$ is the average interaction energy exerted on the surficial water molecules, $E_{ww}$ is the energy provided by the neighboring water molecules; $E_{sub}(q)$ represents the interaction energy from the substrate, mainly provided by the electrical charge *q* assigned on the substrate. We find that the variation of the evaporation flux is the result of a combination of the geometric and the energetic factors.

With respect to molecular dynamics simulation, $P_{geo}(\theta)$ is defined as the ratio of the number of surficial water molecules $N_{surf}$ to the total number of all condensed water molecules N. Since the number of evaporated water molecules in the whole simulation is negligible to the total number of the water adhered on the substrate, N can be taken as a constant:

$$P_{geo}(\theta) = \frac{N_{surf}(\theta)}{N} \tag{2}$$

To estimate how the surface wettability affects $P_{geo}(\theta)$, we presume that the sessile water droplet is a perfect spherical cap. The volume of the outermost molecule layer can be simplified as the upper surface area of the sessile droplet multiplies the thickness of the layer, which is the length of a water molecule. Then $P_{geo}(\theta)$ can be denoted as:

$$P_{geo}(\theta) = \frac{S(\theta) D_{water}}{V_{total}} \tag{3}$$

Here, $D_{water}$ is the diameter of a water molecule (roughly 0.3 nm), $V_{total}$ is the total volume of the droplet, both of which are constant, and $S(\theta)$ is the area of the gas-liquid interface of a sessile water droplet.

$S(\theta)$ can be expressed as a function of contact angle $\theta$:

$$S(\theta) = 2\pi R(\theta)^2 (1 - \cos\theta) \tag{4}$$

where $R(\theta)$ is the radius of curvature of the spherical cap shaped sessile water droplet. Since the total number of the water molecules in the droplet is taken as a constant in the simulation, the total volume of the droplet can also be taken as a constant:

$$V_{total} = \left(\frac{2}{3} + \frac{\cos^3\theta}{3} - \cos\theta\right) \pi R(\theta)^3 \tag{5}$$

According to formula (3), (4) and (5) we can rewrite the geometrical factor $P_{geo}(\theta)$ as a function of the contact angle $\theta$:

$$P_{geo}(\theta) = P_0 \left(\frac{2}{3} + \frac{\cos^3\theta}{3} - \cos\theta\right)^{-\frac{2}{3}} (1 - \cos\theta) \tag{6}$$

where:

$$P_0 = 2D_{water}\left(\frac{\pi}{V_{total}}\right)^{\frac{1}{3}} \tag{7}$$

is constant in simulation.

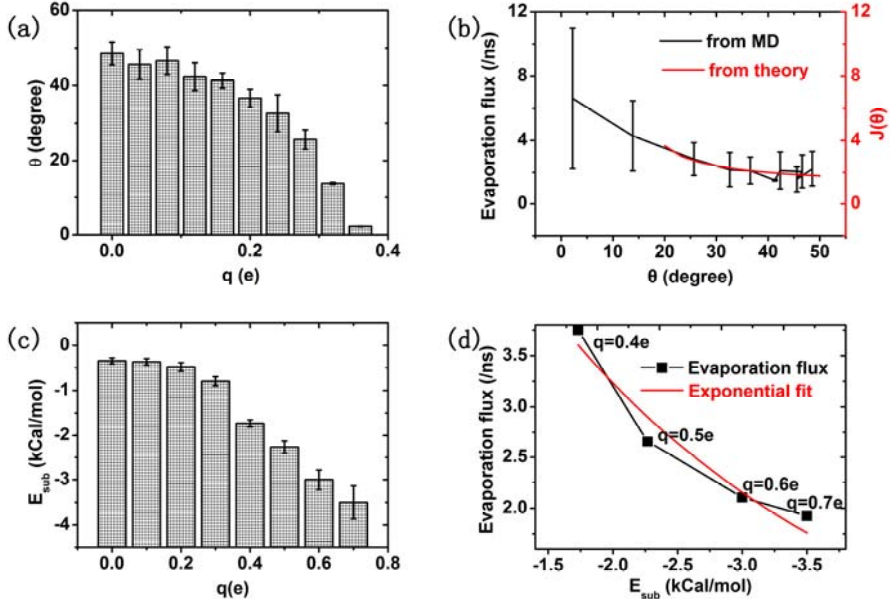

Figure 3 (a) The contact angle θ of the water droplet with different assigned charge $q$. (b) The evaporation flux (black curve with error bar) for $q < 0.4$ e with different contact angle and the theoretical fit $J(\theta)$ (red curve). (c) The interaction energy exerted on the outermost-layer water by the substrate ($E_{sub}$) with different assigned charge $q$. (d) The Evaporation flux when $q \geq 0.4$ e (black line+point) with different $E_{sub}$ and the Exponential fitted (red curve).

The relationship between the assigned charge $q$ and the contact angle of the water droplet is shown in Figure 3a. To demonstrate the continual tunable wettability of the substrate, we sampled the assigned charge $q$ every 0.04 e and the results of the contact angle θ were the statistical average of 4 equilibrium simulations. We find that the contact angle of the water droplet θ decreases as $q$ increases and reaches 0° when $q = 0.4$ e.

When $q < 0.4$ e, most of the surficial water molecules are relatively far from the substrate. According to Figure 3c, the energy $E_{sub}$ provided by the substrate does not change much, and is negligible if compared to the $E_{sub}$ of $q \geq 0.4$ e situation. At the same time, the $E_{ww}$ provided by the neighboring water molecules almost keeps constant in simulation. Hence, for $q < 0.4$ e, the escape probability of a surficial water molecule $P_{ener}(E)$ is nearly a constant.

Thus, for $q < 0.4$ e, the evaporation flux $J(q)$ in formula (1) can be rewritten as below:

$$J(\theta) = J_0 P_{geo}(\theta) \tag{8}$$

As shown in Figure 3b, the variation of the evaporation flux can be perfectly fitted to $J(\theta)$ with $J_0 = 1.24 ns^{-1}$ in the range of $20° < \theta < 50°$ ($\theta = 50°$ corresponds to $q = 0$ e) when the shape of the water droplet still keeps a sessile spherical cap. However, for $\theta < 20°$, the shape of a sessile spherical cap is no longer observed under such situation. Since the expression of $J(\theta)$ is based on the assumption that the water droplet is a perfect sessile spherical cap, formula (8) cannot well fit the variation of the evaporation flux in this region.

For $q \geq 0.4$ e, the adhered water forms a flat single-layer molecule sheet with only a few water molecules overlapping upon it, and the shape of the tiny water aggregation does not change much with different $q$. According to the definition of $P_{geo}(\theta)$, all the water molecules are on the surface layer now, $N_{surf} = N$, therefore $P_{geo}(\theta) = 1$. The formula (1) can be presented as below:

$$J(q) \propto P_{ener}(E_{sub}(q)) \tag{9}$$

$E_{sub}(q)$ is the interaction energy from the substrate, mainly provided by the electrical charge $q$ assigned on the substrate. According to the thermal dynamics, for the system under the NVT ensemble, the probability for a free molecule to gain kinetic energy more than $E_0$ is proportional to $\exp\left(-\dfrac{E_0}{k_B T}\right)$[43,46]. In our simulation, the evaporation flux decreases almost exponentially with respect to $E_{sub}$ as shown on Figure 3d for $q \geq 0.4$ e, which indicates that when the water spreads smoothly on the substrate, the attraction from the substrate becomes the main factor that impacts the water evaporation.

## Conclusion:

The effect of the surface wettability of the substrate on the evaporation of the nanoscale water aggregation adhered on it is the competition of two different mechanism: one is the surface wettability influences the shape of liquid-gas surface of the water aggregation; the other is the attraction provided by the substrate of different wettability changes the escape probability of surficial water molecules. When the surface wettability of the substrate is not high enough, the water molecules accumulate into the form of a sessile spherical cap. The predominant factor that affects the evaporation flux is the geometry shape of the water congregation. The surface area of the droplet is the main factor that influences the evaporation flux. Since the surface area of the droplet increases as the contact angle decreases, the evaporation flux increases as the surface wettability increases. Meanwhile when the surface wettability of the substrate is high enough, the water molecules spread to a monolayer and the geometric factor no longer affects much because the form of the water congregation almost remains the same. In this situation, when the surface wettability increases, the energy barrier provided by the substrate that prevents the water molecules from running off becomes stronger, which cause the decreasing of the evaporation flux.

This study of the effect of the surface wettability on the evaporation of nanoscale water aggregation may enlighten the study of multiple natural evaporation processes that related to the ultrathin water aggregation, which is very important for biological and environmental science.

## AUTHOR INFORMATION:

Corresponding Author:

*To whom correspondence should be addressed E-mail: wanrongzheng@sinap.ac.cn.

## ACKNOWLEDGEMENT:

We gratefully acknowledge Prof. Xinguang Zhu in Shanghai Institutes for Biological Sciences, Prof. Yi Gao, Dr. Chunlei Wang and Dr. Wenpeng Qi for their helpful discussions. This work was supported by This work was supported by NNSFC (10825520, 11105088), SLADP (B111), IPSHMEC (11YZ20), the Knowledge Innovation Program of the Chinese Academy of Sciences, and Shanghai Supercomputer Center of China.

**TOC:**

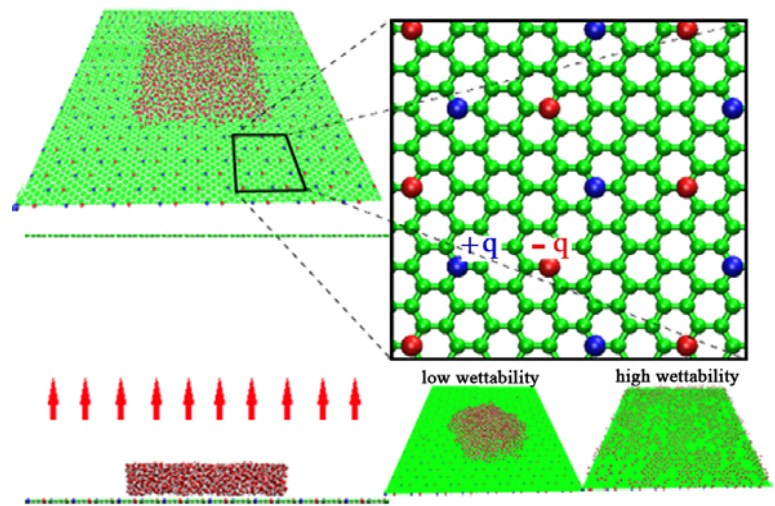